\documentstyle[11pt,epsf]{article}
\catcode`@=11
\def\seceqaa{\@addtoreset{equation}{section}
\def\theequation{A\arabic{equation}}}
\catcode`@=12
\begin{document}
\hfill hep-th/0205293\\

\hfill MRI-O-020501\\
\begin{center}
{\Large \bf
On the Exact Evaluation of the Membrane Instanton Superpotential in
$M$-Theory 
on $G_2$-Holonomy Manifolds} 
\vskip 0.1in
{Aalok Misra\\
Harish Chandra Research Institute,\\
Jhunsi, Allahabad - 211 019, India\\email: aalok@mri.ernet.in}
\vskip 0.5 true in
\end{center}

\begin{abstract}
Following the work of \cite{Ovrutetal} on the exact evaluation of the
nonperturbative contribution to the superpotential from open-membrane instanton in
Heterotic $M$-Theory, we evaluate systematically the contribution to the 
superpotential of a membrane instanton obtained by 
wrapping of a single $M2$ brane, once, on an isolated 
supersymmetric 3-cycle in a $G_2$-holonomy manifold. We
then try to relate the results obtained to those sketched out in
\cite{harvmoore}. We also do a heat-kernel asymptotics analysis 
to see whether one gets similar UV-divergent 
terms for (one or both of) the bosonic
and fermionic determinants indicative of  (partial) cancellation among them. 
The answer is in the affirmative, as expected by the supersymmetry
of the starting membrane action. This work is a first step, both,
in extending the work of \cite{harvmoore} to the 
evaluation of nonperturbative superpotentials for non-rigid
supersymmetric 3-cycle wrappings, and in understanding the
large $N$ Chern-Simons/closed type-$A$ topological string theory 
duality of \cite{GopVafa} from $M$ theory point of view.
\end{abstract}

\section{Introduction}

String and $M$ theories on manifolds with $G_2$ and $Spin(7)$ holonomies 
have become an active area of research, after construction  of explicit
examples of such manifolds by Joyce\cite{joyce}. Some explicit metrics
of noncompact manifolds with the above-mentioned exceptional holonomy groups
have been constructed by  Brandhuber et al\cite{gubseretal} and Cvectic et al
\cite{cveticetal}.

Gopakumar and Vafa in \cite{GopVafa}, had conjectured that similar to the large $N$
Chern-Simons/open topological string theory duality of Witten, large
$N$ Chern-Simons on $S^3$ is dual to closed type-A topological string
theory on an $S^2$-resolved conifold geometry. This conjecture was verified
for arbitrary genus $g$ and arbitrary t'Hooft coupling.
This duality was embedded by Vafa in type IIA to conjecture
the following duality: type IIA with $N$ $D6$ branes wrapping the $S^3$ in
the $S^3$ resolved conifold, is dual to type IIA with the $D6$ branes being
replaced by $RR$ flux through $O(-1)+O(-1)$ line bundle over ${\bf CP^1}$.
The latter (duality) was proven by uplifting it to M theory on a $G_2$
holonomy manifold by Atiyah, Maldacena and Vafa\cite{amv}.  
The $G_2$ holonomy manifold that was considered by Atiyah et al in \cite{amv} 
was a spin bundle over $S^3$ with the topology of $R^4\times S^3$. 
The volume of $S^3$,  gets complexified to $V_M$ with the imaginary 
part given by the flux of the 3-form of $M$ theory through $S^3$. Further,
it is argued that $M$ theory on the above-mentioned $G_2$-holonomy manifold,
modded out by a group $G$, with complex volume $-V_M$ is given by $M$ theory
modded out by a group $G^\prime$, related by a ${\bf Z}_2$ outer
automorphism to the group $G$, and complex volume $V_M$. Then, modding out the
above-mentioned $G_2$-holonomy manifold by two ${\bf Z}_2$ actions, the two
sides of Vafa's type IIA dualities were obtained: the
one with fixed points yielding type IIA theory with $D6$ branes
wrapping $S^3$, and the one without fixed points yielding type IIA theory
with $RR$ flux through $S^2$. The $M$ theory lifts of both sides are
related by a $V_M\leftrightarrow-V_M$ ``flop".

It will be interesting to be able to lift 
the above-mentioned Gopakumar-Vafa duality to $M$ theory on a $G_2$-holonomy
manifold. As the type-A topological string  theory's partition function
receives contributions only from holomorphic maps from the world-sheet
to the target space, and apart from constant maps, instantons fit the bill,
as a first step we should look at obtaining the superpotential contribution
of multiple wrappings of $M2$ branes on supersymmetric 3-cycles in
a suitable $G_2$-holonomy manifold(membrane instantons). A sketch of the
result anticipated for a single $M2$ brane wrapping an isolated
supersymmetric 3-cycle, was given by Harvey-Moore. In this work, we have worked
out the exact expression for the same, using techniques developed in
\cite{Ovrutetal} on evaluation of  the nonperturbative contribution to the
superpotential of open membrane instantons obtained by wrapping the $M2$
brane on an interval [0,1] times (thus converting the problem 
to that of a heterotic string wrapping) 
a holomorphic curve in a Calabi-Yau three-fold.

The paper is organized as follows. In Section {\bf 2}, we spell out the details
of the calculation and obtain the exact form of the nonperturbative contribution
to the superpotential of a single $M2$ brane wrapping an isolated supersymmetric
cycle embedded in the $G_2$-holonomy manifold. In Section {\bf 3}, we compare
our result with the one sketched  out in \cite{harvmoore}. 
Section {\bf 4} has a discussion on heat kernel asymptotics related
to looking for possible cancellations among the bosonic and fermionic
determinants. 
Section {\bf 5} has the conclusion and  some speculative
remarks about connecting this $M$ theory result (including its possible extension
to multiple-covering of wrapping of the $M2$ brane on supersymmetric cycles 
in the $G_2$-holonomy manifold) to the Schwinger 1-loop calculation and the
large $N$ limit of the Chern-Simons partition function of \cite{GopVafaMth1} to get
a step closer in uplifting the Gopakumar/Vafa duality to $M$ theory.

\section{Evaluation of the membrane instanton contribution to the superpotential}

As given in \cite{harvmoore}, the Euclidean action for an $M2$ brane is
given
by the following Bergshoeff, Sezgin, Townsend action:
\begin{equation}
\label{eq:EucM2}
{\cal S}_\Sigma=\int d^3z\Biggl[{\sqrt{g}\over l_{11}^3}
-{i\over 3!}\epsilon^{ijk}
\partial_i{\bf Z}^M\partial_j{\bf Z}^N\partial_k{\bf Z}^PC_{MNP}
(X(s),\Theta(s))\Biggr],
\end{equation}
where ${\bf Z}$ is the map of the $M2$ brane world-volume to the
the $D=11$ target space $M_{11}$, both being regarded as supermanifolds.
The $g$ in (\ref{eq:EucM2}), is defined as:
\begin{equation}
\label{eq:gdef}
g_{ij}=\partial_i{\bf Z}^M\partial_j{\bf Z}^N{\bf E}^A_M{\bf
E}^B_N\eta_{AB},
\end{equation}
where ${\bf E}^A_M$ is the supervielbein, given in \cite{harvmoore}.
$X(s)$ and $\Theta(s)$ are the bosonic and fermionic coordinates
of ${\bf Z}$. After using the static gauge and 
$\kappa$-symmetry fixing, the physical degrees of freedom, are given
by $y^{m^{\prime\prime}}$, the section of the normal bundle to
the $M2$-brane world volume, and $\Theta(s)$, section of the 
spinor bundle tensor product: $S(T\Sigma)\otimes S^-(N)$, where
the $-$ is the negative $Spin(8)$ chirality, as under an orthogonal 
decomposition of $TM_{11}|_\Sigma$ in terms of tangent and normal 
bundles, the structure group $Spin(11)$ decomposes into
$Spin(3)\times Spin(8)$.

The action in (\ref{eq:EucM2}) needs to be expanded up to $O(\Theta^2)$,
and the expression is (one has to be careful that in Euclidean $D=11$, one does
not have a Majorana-Weyl spinor or a Majorana spinor) given as:
\begin{eqnarray}
\label{eq:actexpTh2}
& & {\cal S}_\Sigma=\int_\Sigma\Biggl[C +{i\over l_{11}^3}vol(h)
+{\sqrt{h_{ij}}\over l_{11}^3}\biggl(h^{ij}
D_iy^{m^{\prime\prime}} D_j y^{n^{\prime\prime}}
h_{m^{\prime\prime}n^{\prime\prime}}-y^{m^{\prime\prime}}
{\cal U}_{m^{\prime\prime}n^{\prime\prime}}y^{n^{\prime\prime}}+O(y^3)
\biggr)\nonumber\\
& & +{i\over l_{11}^3}\sqrt{h_{ij}}{1\over2}({\bar\Psi}_MV^M 
- {\bar V}^M\Psi_M)+2{\sqrt{h_{ij}}\over l_{11}^3}h^{ij}
{\bar\Theta}\Gamma_iD_j\Theta + O(\Theta^3)\Biggr],
\end{eqnarray}
where we follow the conventions of \cite{harvmoore}: $V_M$ being
the gravitino vertex operator, $\Psi$ being the gravitino field that
enters via the supervielbein ${\bf E}^A_M$, ${\cal U}$ is a mass
matrix defined in terms of the Riemann curvature tensor and the
second fundamental form (See (\ref{eq:Udieuf})).

After $\kappa$-symmetry fixing, like \cite{harvmoore}, we set
$\Theta_2^{A\stackrel{.}{a}}(s)$, i.e., the positive $Spin(8)$-chirality to zero,
and following \cite{Ovrutetal}, will refer to $\Theta_1^{Aa}(s)$ as
$\theta$.

The Kaluza-Klein reduction of the $D=11$ gravitino is given by:
\begin{eqnarray}
\label{eq:KKgrav}
& & dx^M\Psi_M=dx^\mu\Psi_\mu+dx^m\Psi_m,\nonumber\\
& & \Psi_\mu(x,y)=\psi_\mu(x)\otimes\vartheta(y),\nonumber\\
& & \Psi_m(x,y)=l_{11}^3\sum_{I=1}^{b_3}
\omega^{(3)}_{I,mnp}(y)\Gamma^{pq}\chi^I(x)\otimes\tilde{\eta}(y),
\end{eqnarray}
where we do not write the terms obtained by expanding in terms of
$\{\omega^{(2)}_{I,mn}\}$, the harmonic 2-forms forming a basis for
$H^2(X_{G_2},{\bf Z})$, as we will be interested in $M2$ branes
wrapping supersymmetric 3-cycles in the $G_2$-holonomy manifold.
For evaluating the nonperturbative contribution to the superpotential,
following \cite{harvmoore}, we will evaluate the fermionic 2-point
function: $\langle\chi^i(x_1^u)\chi^j(x_2^u)\rangle$ (where $x_{1,2}$ are the
${\bf R}^4$ coordinates and u [and later also $v$]$\equiv 7,8,9,10$
is [are] used to index these coordinates), 
and drop the interaction terms in the $D=4,{\cal N}=1$ supergravity action. 
The corresponding mass
term in the supergravity action appears as $\partial_i\partial_j W$,
where the derivatives are evaluated w.r.t. the complex scalar
obtained by the Kaluza-Klein reduction of $C+{i\over l_{11}^3}\Phi$ using 
harmonic three forms forming a basis for $H^3(X_{G_2},{\bf R})$.
One then integrates twice to get the expression for the superpotential
from the 2-point function.

The bosonic zero modes are the four bosonic coordinates that specify
the position of the supersymmetric 3-cycle, and will be denoted
by $x_0^{7,8,9,10}\equiv x^u_0$. The fermionic zero modes come from
the fact that for every $\theta_0$ that is the solution to the fermionic
equation of motion, one can always shift $\theta_0$ to
$\theta_0+\theta^\prime$
, where $D_i\theta^\prime=0$. This $\theta^\prime=\vartheta\otimes\eta$
where $\vartheta$ is a $D=4$ Weyl spinor, and $\eta$ is a covariantly
constant spinor on the $G_2$-holonomy manifold.

After expanding the $M2$-brane action in fluctuations about  solutions to
the bosonic and fermionic equations of motion, one gets that:
\begin{equation}
\label{eq:expfl1}
{\cal S}|_\Sigma={\cal S}^y_0+{\cal S}^\theta_0+{\cal S}^y_2
+{\cal S}^\theta_2,
\end{equation}
where 
\begin{eqnarray}
\label{eq:expfl2}
& & {\cal S}^y_0\equiv {\cal S}_\Sigma|_{y_0,\theta_0};\nonumber\\
& & {\cal S}^\theta_0\equiv {\cal S}_\Sigma^\theta
+{\cal S}^{\theta^2}_\Sigma|_{y_0,\theta_0};\nonumber\\
& & {\cal S}^y_2
\equiv {\delta^2{\cal S}_\Sigma\over\delta y^2}|_{y_0,\theta_0=0}
(\delta y)^2;\nonumber\\
& & {\cal S}^\theta_2
\equiv {\delta^2{\cal S}\over\delta\theta^2}|_{y_0,\theta_0=0}
(\delta\theta)^2.
\end{eqnarray}
In (\ref{eq:expfl2}), following \cite{Ovrutetal}, we consider
classical values of coefficients of $(\delta y)^2,(\delta\theta)^2$ terms,
as fluctuations are considered to be of ${\cal O}(\sqrt{\alpha^\prime})$.

Now, 
\begin{eqnarray}
\label{eq:2ptdef}
& & \langle \chi^i(x_1^u)\chi^j(x_2^u)\rangle=\nonumber\\
& & \int {\cal D}\chi e^{K.E\ of\ \chi}\chi^i(x)\chi^j(x)
\int d^4x_0 e^{-{\cal S}_0^y}\nonumber\\
& & \times\int d\vartheta^1d\vartheta^2 e^{-{\cal S}^\theta_0}
\int{\cal D}\delta y^{m^{\prime\prime}}e^{-{\cal S}^y_2}
\int{\cal D}\delta{\bar\theta}{\cal D}\delta\theta
e^{-{\cal S}^\theta_2}.
\end{eqnarray}

We now evaluate the various integrals that appear in (\ref{eq:2ptdef}) above
starting with $\int d^4x e^{-{\cal S}_0^y}$:
\begin{equation}
\label{eq:bzmint}
\int d^4x_0 e^{-{\cal S}_0^y}=\int d^4x_0 e^{[iC-{1\over l_{11}^3}vol(h)]}.
\end{equation}

Using the 11-dimensional Euclidean representation of the gamma matrices
as given in \cite{harvmoore},
\begin{eqnarray}
\label{eq:D=11Gammadefs}
& & \Gamma^{1,2,3}=\sigma^{1,2,3}\otimes\left(\begin{array}{cc}\\
-{\bf 1}_8 & 0\\
0 & {\bf 1}_8\\
\end{array}\right)\nonumber\\
& & \Gamma^{4,5,6,7,8,9,10}={\bf 1}_2\otimes
\left(\begin{array}{cc}\\
0 & \gamma^{1,2,3,4,5,6,7}\\
-\gamma^{1,2,3,4,5,6,7} & 0 \\
\end{array}\right)\nonumber\\
& & \Gamma^{11}={\bf 1}_2\otimes\left(\begin{array}{cc}
0 & \gamma^8 \\
\gamma^8 & 0 \\
\end{array}\right)
\end{eqnarray}
where $\gamma^{1,2,3,4,...,8}\in Cl(8)$.
 and that on-shell,
\begin{equation}
{\cal S}^\theta_0+{\cal S}^{\theta^2}_0|_\Sigma={i\over
2l_{11}^3}\int_\Sigma
\sqrt{h_{ij}}{\bar\Psi}_MV_M d^3s,
\end{equation}
where using $\partial_i x_0^u=0$, and using $U$ to denote coordinates
on the $G_2$-holonomy manifold, $V^U=h_{ij}\partial_i y_0^U\partial_j y^V
\gamma_V\theta_0+{i\over2}\epsilon^{ijk}\partial_iy_0^U
\partial_jy_0^V\partial_k y_0^W\Gamma_{VW}\theta_0$,
\begin{equation}
\label{eq:linth}
\int d\vartheta_1 d\vartheta_2e^{{i\over 2l_{11}^3}\sum_{I=1}^{b_3}
\sum_{\alpha=1}^2\sum_{i=1}^8
({\bar\chi}(x)\sigma^{(i)})_\alpha\vartheta_\alpha\omega_I^{(i)}}
=-{1\over 4l_{11}^3}\sum_{I=1}^{b_3}\sum_{i<j=1}^8\omega_I^{(i)}
\omega_I^{(j)}({\bar\chi}\sigma^{(i)})_1({\bar\chi}\sigma^{(j)})_2,
\end{equation}
where,
\begin{eqnarray}
\label{eq:defs}
& & \sigma^{(1)}\equiv \sigma^{V_2W_2}\sigma^{V_1},\ V_{1,2},W_2\in1,2,3,
\nonumber\\
& & \omega_I^{(1)}\equiv\int_\Sigma d^3s\omega_{I,UV_2W_2}(y)\sqrt{h_{ij}}
h_{ij}\partial_iy_0^U\partial_jy_0^{V_1}{\bar{\tilde{\eta}}}(y)
\left(\begin{array}{cc}\\
-{\bf 1}_8 & 0\\
0 & {\bf 1}_8 \\
\end{array}\right)\eta(y);\nonumber\\
& & \sigma^{(2)}\equiv \sigma^{V_2W_2}\sigma^{W_1V_1},\ V_{1,2},W_2\in1,2,3,
\nonumber\\
& & \omega_I^{(2)}\equiv\int_\Sigma d^3s\omega_{I,UV_2W_2}(y)\sqrt{h_{ij}}
\epsilon^{ijk}\partial_iy_0^U\partial_jy_0^{V_1}\partial_ky_0^{W_1}
{\bar{\tilde{\eta}}}(y)\gamma^{W_1V_1}\otimes{\bf 1}_2\eta(y);\nonumber\\
& & \sigma^{(3)}\equiv \sigma^{V_2W_2}\ V_1\in4,...,7,\ V_2,W_2\in1,2,3,
\nonumber\\
& & \omega_I^{(3)}\equiv\int_\Sigma d^3s\omega_{I,UV_2W_2}(y)\sqrt{h_{ij}}h_{ij}
\partial_iy_0^U\partial_jy_0^{V_1}{\bar{\tilde{\eta}}}(y)
\left(\begin{array}{cc}\\
0 & \gamma^{V_1}\\
-\gamma^{V_1} & 0 \\
\end{array}\right)\eta(y);\nonumber\\
& & \sigma^{(4)}\equiv \sigma^{V_2W_2}\sigma^{V_1},\ V_1,W_1\in4,...,7,\ 
V_2,W_2\in1,2,3
\nonumber\\
& & \omega_I^{(4)}\equiv\int_\Sigma d^3s\omega_{I,UV_2W_2}(y)\sqrt{h_{ij}}
\epsilon^{ijk}
\gamma^{W_1V_1}\otimes{\bf 1}_2
\partial_iy_0^U\partial_jy_0^{V_1}\partial_ky_0^{W_1}{\bar{\tilde{\eta}}}(y)
\eta(y);\nonumber\\
& & \sigma^{(5)}\equiv \sigma^{V_1},\ V_1\in1,2,3,\ V_2,W_2\in4,...,7
\nonumber\\
& & \omega_I^{(5)}\equiv\int_\Sigma d^3s\omega_{I,UV_2W_2}(y)\sqrt{h_{ij}}
h_{ij}\partial_iy_0^U\partial_jy_0^{V_1}{\bar{\tilde{\eta}}}(y)
\gamma^{V_2W_2}\otimes{\bf 1}_2
\left(\begin{array}{cc}\\
-{\bf 1}_8 & 0\\
0 & {\bf 1}_8 \\
\end{array}\right)\eta(y);\nonumber\\
& & \sigma^{(6)}\equiv \sigma^{V_1W_1},\ V_1,W_1\in1,2,3,\ V_2W_2\in4,...,7,
\nonumber\\
& & \omega_I^{(1)}\equiv\int_\Sigma d^3s\omega_{I,UV_2W_2}(y)\sqrt{h_{ij}}
\epsilon^{ijk}
\partial_iy_0^U\partial_jy_0^{V_1}\partial_ky_0^{W_1}{\bar{\tilde{\eta}}}(y)
\gamma^{V_2W_2}\otimes{\bf 1}_2\eta(y);\nonumber\\
& & \sigma^{(7)}\equiv {\bf 1},\ V_{1,2}W_2\in4,...,7,
\nonumber\\
& & \omega_I^{(7)}\equiv\int_\Sigma d^3s\omega_{I,UV_2W_2}(y)\sqrt{h_{ij}}
h_{ij}
\partial_iy_0^U\partial_jy_0^{V_1}{\bar{\tilde{\eta}}}(y)\gamma^{V_2W_2}
\otimes{\bf 1}_2
\left(\begin{array}{cc}\\
0 & \gamma^{V_1}\\
-\gamma^{V_1} & 0 \\
\end{array}\right)\eta(y);\nonumber\\
& & \sigma^{(8)}\equiv {\bf 1},\ V_{1,2},W_{1,2}\in4,...,7,
\nonumber\\
& & \omega_I^{(1)}\equiv\int_\Sigma d^3s\omega_{I,UV_2W_2}(y)\sqrt{h_{ij}}
\epsilon^{ijk}
\partial_iy_0^U\partial_jy_0^{V_1}\partial_ky_0^{W_1}{\bar{\tilde{\eta}}}(y)
\gamma^{V_2W_2}\otimes{\bf 1}_2\gamma^{V_1W_1}\eta(y);
\end{eqnarray}
where $\eta(y)$ is the 7-dimensional component of $\theta$ and
$\tilde{\eta}(y)$ is the 7-dimensional component of $\Psi_U$.
We follow the following notations for coordinates:
$u,v$ are ${\bf R}^4$ coordinates, $\hat{U},\hat{V}$ 
are $G_2$-holonomy manifold
coordinates that are orthogonal to the $M2$ world volume (that wraps
a supersymmetric 3-cycle embedded in the $G_2$-holonomy manifold),
and $U,V$ are $G_2$-holonomy manifold coordinates. The 
tangent/curved space coordinates for $\Sigma$
are represented by $a^\prime/m^\prime$ and those for
$X_{G_2}\times{\bf R}^4$ are represented by 
$a^{\prime\prime}/m^{\prime\prime}$.

We now come to the evaluation of ${\cal S}^\theta_2|_{y_0,\theta_0=0}$.
Using the equality of the two $O((\delta)\Theta^2)$ terms 
in the action of Harvey
and Moore, and arguments similar to the ones in \cite{Ovrutetal}, one can show
that
\begin{eqnarray}
\label{eq:th2exps}
& & {\cal S}^\theta_2|_{y_0,\theta_0=0}
={2i\over l_{11}^3}\int d^3s\sqrt{H_{ij}}\delta{\bar\Theta}\Gamma_jD_i\delta\Theta
={2i\over l_{11}^3}\int d^3s\sqrt{h_{Ij}}h^{ij}\partial_jX^M\delta{\bar\Theta}\Gamma_MD_i\delta\Theta\nonumber\\
& & ={2i\over l_{11}^3}\int d^3s\sqrt{h_{ij}}h^{ij}\partial_jX^M[\delta{\bar\Theta}\Gamma_M\partial_i\delta\Theta+{\bar\Theta}\Gamma_M\omega_N^{AB}\Gamma_{AB}\delta\Theta\partial_i
X^N]\nonumber\\
& & ={2i\over l_{11}^3}
\int d^3s\sqrt{h_{ij}}\biggl[
h^{ij}\delta^{m^\prime}_j
e^{a^\prime}_{m^\prime}\delta{\bar\Theta}(\Gamma_{a^\prime}\partial_i+\omega^{AB}_i
\Gamma_{a^\prime}\Gamma_{AB})\delta\Theta\nonumber\\
& & +h^{ij}\partial_j
y^{\prime\prime}e^{a^{\prime\prime}}_{m^{\prime\prime}}\delta{\bar\Theta}(
\Gamma_{a^{\prime\prime}}\partial_i+\omega^{AB}_i\Gamma_{m^{\prime\prime}}
\Gamma_{AB})\delta\Theta\biggr],
\end{eqnarray}
from where one sees that one needs to evaluate the following bilinears:
$\delta{\bar\Theta}\Gamma_{a^\prime}\partial_i\delta\Theta$,
$\delta{\bar\Theta}\Gamma_{a^{\prime\prime}}\partial_i\delta\Theta$,
$\delta{\bar\Theta}\Gamma_{a^\prime}\Gamma_{AB}\delta\Theta$, and
$\delta{\bar\Theta}\Gamma_{m^{\prime\prime}}\Gamma_{AB}\delta\Theta$.
One can then show (using that ${\bar\psi}=\psi^\dagger$ in Euclidean space:
\begin{eqnarray}
\label{eq:non0bils}
& & \delta{\bar\Theta}\Gamma_3\partial_i\delta\Theta=-\delta\theta^\dagger\sigma_3\otimes
{\bf 1}_8\delta\Theta;\nonumber\\
& & \delta{\bar\Theta}\Gamma_{a^{\prime\prime}}\partial_i\delta\Theta=i\delta\theta^\dagger
\sigma^2\otimes\gamma_{a^{\prime\prime}}\delta\theta;\nonumber\\
& & \delta{\bar\Theta}\Gamma_1\Gamma_{23}\delta\Theta=-i\delta\theta^\dagger\sigma^3
\otimes{\bf 1}_8\delta\theta;\nonumber\\
& & \delta{\bar\Theta}\Gamma_1\Gamma_{12}\delta\Theta=\delta{\bar\Theta}\Gamma_1
\Gamma_{13}\delta\Theta=0;\nonumber\\
& & \delta{\bar\Theta}\Gamma_2\Gamma_{23}\delta\Theta=\delta{\bar\Theta}\Gamma_2
\Gamma_{23}\delta\Theta=-\delta\theta^\dagger\sigma^3\otimes{\bf 1}_8\delta\theta;
\nonumber\\
&& \delta{\bar\Theta}\Gamma_{a^\prime}\Gamma_{b^{\prime\prime}}\delta\theta
=\delta_{a^\prime}^3\delta\theta^\dagger\sigma^3\otimes\gamma^{b^{\prime\prime}}
\gamma^{c^{\prime\prime}}\delta\theta;\nonumber\\
& & \delta{\bar\Theta}\Gamma_{a^{\prime\prime}}\Gamma_{b^{\prime\prime}c^{\prime\prime}}
\delta\Theta=-i\delta\theta^\dagger\sigma^2\otimes\gamma^{a^{\prime\prime}}
\gamma^{b^{\prime\prime}}\gamma^{c^{\prime\prime}}\delta\theta.
\end{eqnarray}
Using (\ref{eq:non0bils}), one gets:
\begin{eqnarray}
\label{eq:th2}
& & {\cal S}^\theta_2|_{y_0,\theta_0=0}\equiv\int_\Sigma d^3s \delta
\theta^\dagger{\cal O}_3\delta\theta,\nonumber\\
& & {\rm where}\ {\cal O}_3\equiv\nonumber\\
& & {2i\over l_{11}^3}\sqrt{h_{ij}}\Biggl[h^{ij}\delta^{m^\prime}_j
\biggl(-e^3_{m^\prime}\sigma^3\otimes{\bf 1}_8\partial_i
-2i[e^1_{m^\prime}\omega^{23}_i+e^2_{m^\prime}\omega^{31}_i+e^3_i\omega^{1
2}_i]
\sigma^3\otimes{\bf 1}_8\nonumber\\
& & -2[e^1_{m^\prime}\omega^{13}_i+e^2_{m^\prime}\omega^{23}_i]
\sigma^3\otimes{\bf 1}_8
+e^3_{m^\prime}{\omega^{b^{\prime\prime}c^{\prime\prime}}\over2}\sigma^3\otimes
\gamma_{b^{\prime}c^{\prime}}\biggr)\nonumber\\
& & +ih^{ij}e^{a^\prime}_{m^\prime}\omega^{b^\prime c^{\prime\prime}}_i
\sigma^2\otimes\gamma_{c^{\prime\prime}}[\delta^{a^\prime b^\prime}
+i\epsilon^{a^\prime b^\prime c^\prime}\delta^{c^\prime}_3]\nonumber\\
& & +ih^{ij}\partial_j y^{m^{\prime\prime}}
e^{a^{\prime\prime}}_{m^{\prime\prime}}[
\sigma^2\otimes\gamma_{a^{\prime\prime}}\partial_i
-(\omega^{b^{\prime\prime}c^{\prime
\prime}}_i\sigma^2{1\over6}
\otimes\gamma_{a^{\prime\prime}b^{\prime\prime}c^{\prime\prime}}
-{1\over2}\omega^{b^\prime c^{\prime\prime}}\delta^{b^\prime}_3
\sigma^3\otimes\gamma_{a^{\prime\prime}c^{\prime\prime}})]\Biggr],\nonumber\\
& & 
\end{eqnarray}
Hence, the integral over the fluctuations in $\theta$ will give a
factor of $\sqrt{det{\cal O}_3}$ in Euclidean space. 

The expression for ${{\cal S}^y_2|_\Sigma}_{y_0,\theta_0=0}$ is identical
to the one given in \cite{Ovrutetal}, and will contribute 
${1\over{\sqrt{det{\cal O}_1det{\cal O}_2}}}$, where ${\cal O}_1$ and
${\cal O}_2$ are as given in the same paper:
\begin{eqnarray}
\label{eq:bosondets}
& & {\cal O}_1\equiv\eta_{uv}\sqrt{g}g^{ij}{\cal D}_i\partial_j\nonumber\\
& & {\cal O}_2\equiv\sqrt{g}(g^{ij}{\cal D}_ih_{\hat{U}\hat{V}}
D_j+{\cal U}_{\hat{U}\hat{V}}).
\end{eqnarray}
The mass matrix ${\cal U}$ is expressed in terms of the curvature
tensor and product of two second fundamental forms.
${\cal D}_i$ is a covariant derivative with indices in the corresponding
spin-connection of the type $(\omega_i)^{m^{\prime\prime}}_{n^{\prime\prime}}$
and $(\omega_i)^{m^\prime}_{n^\prime}$, and $D_i$ is a covariant derivative
with  corresponding spin connection indices of the former type.

Hence, modulo supergravity determinants, and the contribution from the
fermionic zero modes, the exact form of the superpotential
contribution coming from a single $M2$ brane wrapping an isolated supersymmetric
cycle of $G_2$-holonomy manifold, is given by:
\begin{equation}
\label{eq:Wfinal}
\Delta W = e^{iC - {1\over l_{11}^3}vol(h)}\sqrt{{det O_3\over
det {\cal O}_1\ det{\cal O}_2}}.
\end{equation}
As in \cite{gubseretal}, we do not bother about 5-brane instantons, as
we assume that there are no supersymmetric 6-cycles in the $G_2$-holonomy
manifold that we consider.

\section{Comparison with Harvey-Moore's paper}

In \cite{harvmoore}, it is argued that for 
an ``associative'' 3-fold $\Sigma$
in the $G_2$-holonomy manifold, the structure group $Spin(8)$ decomposes
into $Spin(4)_{\bf R^4}\times Spin(4)_{X_{G_2}\setminus\Sigma}$. After
gauge-fixing under $\kappa$-symmetry, 
\begin{equation}
\label{eq:thcomps}
\Theta=\biggl((\Theta_{--})^{AY}_\alpha, 
(\Theta_{++})^{\stackrel{\cdot}{Y}}_{\stackrel{\cdot}{\alpha} A};
0,0\biggr),
\end{equation}
where $A,\stackrel{(\cdot)}{\alpha},\stackrel{(\cdot)}{Y}$ 
are the $Spin(3), Spin(4)_{\bf R^4},
 Spin(4)_{X_{G_2}\setminus\Sigma}$ indices respectively. The $G_2$ structure allows
one to trade off $(\Theta_{--})^{AY}_\alpha$  for fermionic 0- and 1-forms:
$\eta,\chi_i$, which together with $y^u\equiv 
y^{\alpha\stackrel{\cdot}{\alpha}}$, form
the Rozansky-Witten(RW) multiplet. Similarly, $(\Theta_{++})^{
\stackrel{\cdot}{Y}}_{\stackrel{\cdot}{\alpha} A}$
gives the Mclean multiplet: $
(y^{A\stackrel{\cdot}{y}},\nu^{\stackrel{\cdot}{Y}}_{\stackrel{\cdot}{\alpha}
 A})$.
The RW model is a $D=3$ topological sigma model on a manifold
embedded in a hyper-K\"{a}hler manifold $X_{4n}$ \cite{RozWitt}. If
$\phi^{M(=1,...,4n)}(x^i)$
are functions from mapping $M$ to $X$, then the RW action is given by:
\begin{eqnarray}
\label{eq:RW}
& & 
\int_\Sigma\sqrt{h_{ij}}\Biggl[{1\over2}h_{MN}\partial_i\phi^M\partial_j\phi^N
h^{ij}+\epsilon_{IJ}h^{ij}\chi^I_iD_j\eta^J\nonumber\\
& & +{1\over2\sqrt{h_{ij}}}\epsilon^{ijk}\biggl(\epsilon_{IJ}
\chi^I_iD_j\chi^J_k+{1\over3}\Omega_{IJKL}\chi_i^I\chi_j^J\chi^L_k\eta^L\biggr) 
\Biggr],
\end{eqnarray}
where $\Omega_{IJKL}=\Omega_{JIKL}=\Omega_{IJLK}$. Then, dropping
the term proportional to $\Omega_{IJKL}$, one sees that the terms in 
(\ref{eq:actexpTh2}),  are very likely to give the RW action in
(\ref{eq:RW}).
In \cite{harvmoore}, $n=1$. 

As the RW and Mclean's multiplets are both contained in 
$\delta\theta$, hence (using the notations of \cite{harvmoore})
$det^\prime(L_-)det^\prime(\rlap/D_E)$ will be
given by $det{\cal O}_3$ - it is
however difficult to disentangle the two contributions.
The relationship
involving the spin connections on the tangent bundle and normal bundle (the anti
self-dual part of the latter) as given in \cite{harvmoore}, can be used to
reduce the number of independent components of the spin connection
and thus simplify (\ref{eq:th2}).  
Further, $(det^\prime\Delta_0)^2|det^\prime(\rlap/D_E)|$ 
should be related to $\sqrt{det{\cal O}_1det{\cal O}_2}$. 
Hence, 
the order of $H_1(\Sigma,{\bf Z})$ must
be expressible in terms of $\sqrt{det {\cal O}_{1,2,3}}$ for
$M2-$brane wrapping a rigid supersymmetric 3-cycle. However, we wish to
emphasize that (\ref{eq:th2}), unlike the corresponding result
of \cite{harvmoore}, 
is equally valid for $M2-$brane wrapping a non-rigid supersymmetric 3-cycle,
as considered in Section 4.

\section{Heat Kernel Asymptotics}

In this section we explore the possibility of cancellations 
between the bosonic and fermionic determinants\footnote{
We thank J.Park for suggesting to look for possible cancellations.}. 
For bosonic determinants $det A_b$, the function that is relevant
is $\zeta(s|A_b)$, and that for fermionic determinants $det A_f$, 
the function that is additionally
relevant is $\eta(s|A_f)$. The integral representation of the former involves
$Tr(e^{-tA_b})$, while that for the latter involves $Tr(Ae^{-tA^2})$
(See \cite{elizalde}):
\begin{eqnarray}
\label{eq:zetaetaMellin}
& & \zeta(s|A_b)={1\over\Gamma(2s)}\int_0^\infty dt t^{s-1}
Tr(e^{-tA_b});\nonumber\\
& & \eta(s|A_f)={1\over\Gamma({s+1\over2})}\int_0^\infty
dt t^{{s+1\over2}}Tr(A_fe^{-tA_F^2}) 
, 
\end{eqnarray}
where to get
the UV-divergent 
contributions, one looks at the $t\rightarrow0$ limit of the two terms.        
To be  more precise (See \cite{Deseretal})\footnote{We are grateful to 
K.Kirsten for bringing \cite{Deseretal} to our attention.},
\begin{eqnarray}
\label{eq:bosfermdets}
& & ln det A_b = -{d\over ds}\zeta(s|A_b)|_{s=0}\nonumber\\
& & =-{d\over ds}\biggl({1\over\Gamma(s)}\int_0^\infty dt t^{s-1}
Tr(e^{-tA_b})\biggr)|_{s=0};\nonumber\\
& & ln det A_f = -{1\over2}{d\over ds}\zeta(s|A_f^2)|_{s=0}
\mp{i\pi\over 2}\eta(s|A_f)|_{s=0} \pm{i\pi\over 2}\zeta(s|A_f^2)|_{s=0}
\nonumber\\
& & = \biggl[-{1\over2}{d\over ds}\pm{i\pi\over2}\biggr]\biggl(
{1\over\Gamma(s)}\int_0^\infty dt t^{s-1}Tr(e^{-tA_f^2})\biggr)|_{s=0}
\mp{i\pi\over2}{1\over\Gamma({s+1\over2})}\int_0^\infty dt t^{{s+1\over2}-1}
Tr(A_f e^{-tA_f^2})|_{s=0},\nonumber\\
& & 
\end{eqnarray}
where the $\mp$ sign in front of $\eta(0)$, a non-local object, represents
an ambiguity in the definition of the determinant. The $\zeta(0|A_f^2)$ term
can be reabsorbed into the contribution of $\zeta^\prime(0|A_f^2)$, and
hence will be dropped below.  
Here $Tr\equiv\int dx\langle x|...|x\rangle\equiv\int dx tr(...)$. The idea
is that if one gets a match in the Seeley - de Witt coefficients for
the bosonic and fermionic determinants, implying equality of 
UV-divergence, this is indicative of a possible complete cancellation. 

The heat kernel expansions for the bosonic and fermionic determinants\cite{Gilkey}
are given by:
\begin{eqnarray}
\label{eq:heatkernexps}
& & tr(e^{-tA_b})=\sum_{n=0}^\infty e_n(x,A_b) t^{{(n-m)\over2}},\nonumber\\
& & tr(A_fe^{-tA_f^2})=\sum_{n=0}^\infty a_n(x,A_f) t^{{(n-m-1)\over2}},
\end{eqnarray}
where for $m$ is the dimensionality of the space-time. For our case, we have a compact
3-manifold, for which $e_{2p+1}=0$ and $a_{2p}=0$. For Laplace-type operators $A_b$,
and Dirac-type operators $A_f$, the non-zero coefficients are determined to be the
following:
\begin{eqnarray}
\label{eq:e_ns}
& & e_0(x,A_b)=(4\pi)^{-{3\over2}}Id,\nonumber\\
& & e_2(x,A_b)=(4\pi)^{-{3\over2}}\biggl[\alpha_1 E+\alpha_2\tau Id\biggr],
\end{eqnarray}
where $\alpha_i$'s are constants, $\tau\equiv R_{ijji}$, and
$Id$ is the identity that figures with the scalar leading symbol in the
Laplace-type operator $A_b$ (See \cite{Gilkey}), and
\begin{equation}
\label{eq:Edieuf}
E\equiv B - G^{ij}(\partial_i\omega_j+\omega_i\omega_j-\omega_k\Gamma^k_{ij}),
\end{equation}
where $B$ and $A_i$ are defined via:
\begin{equation}
\label{eq:A^iBdieufs}
A_b\equiv-(G^{ij}Id\partial_i\partial_j+A^i\partial_i+B),
\end{equation}
and
\begin{equation}
\label{eq:omegadieuf}
\omega_i={G_{ij}(a^j+G^{kl}\Gamma^j_{kl}Id)\over 2}.
\end{equation}
For the bosonic operators ${\cal O}_1$ and ${\cal O}_2$, the following are the relevant
quantities:
\begin{eqnarray}
\label{eq:O12qts}
& & {\cal O}_1:\nonumber\\
& & G^{ij}\equiv -\eta_{uv}\sqrt{g}g^{ij};\nonumber\\
& & A^i\equiv -\eta_{uv}\sqrt{g}g^{ij}\omega^{\cal D}_j;\nonumber\\
& & B=0;\nonumber\\
& & \omega_i\equiv\omega^{\cal D}_i+{\sqrt{g}g^{kl}\over2}\biggl[\partial_k\biggl({g_{li}\over
\sqrt{g}}\biggr)+\partial_l\biggl({g_{ki}\over\sqrt{g}}\biggr)-\partial_i\biggl({g_{kl}\over
\sqrt{g}}\biggr)\biggr],
\nonumber\\
& & {\cal O}_2:\nonumber\\
& & G^{ij}\equiv-\sqrt{g}h_{UV}g^{ij};\nonumber\\
& & A^i\equiv-\sqrt{g}g^{ij}\biggl[\partial_jh_{UV}
+h_{UV}\omega^{\cal D}_j+\omega^{\cal D}_jh_{UV}\biggr];\nonumber\\
& & B
\equiv-\sqrt{g}g^{ij}\biggl[\partial_ih_{UV}\omega^D_j
+h_{UV}\partial_i\omega_j^{\cal D}
+h_{UV}\omega^D_i\omega^{\cal D}_j\biggr]-2\sqrt{g}{\cal U}_{UV};\nonumber\\
& & \omega_i\equiv{g_{ij}\over\sqrt{g}h_{UV}}\biggl[\sqrt{g}g^{ij}
\biggl(\partial_ih_{UV}+h_{UV}\omega^D_i+\omega^{\cal D}_ih_{UV}\biggr)
\nonumber\\
& & 
+{gh^2_{UV}g^{kl}g^{jm}\over2}
\Biggl[\partial_i\biggl({g_{lm}\over\sqrt{g}h_{UV}}\biggr)
+\partial_l\biggl({g_{km}\over\sqrt{g}h_{UV}}\biggr)
-\partial_m\biggl({g_{kl}\over\sqrt{g}h_{UV}}\biggr)\biggr]
\Biggr],
\end{eqnarray}
where
\begin{eqnarray}
\label{eq:DcalDdieufs}
& & D_i\equiv\partial_i+\omega^D_i;\nonumber\\
& & {\cal D}_i\equiv\partial_i+\omega^{\cal D}_i.
\end{eqnarray}
To actually evaluate $e_0$ and $e_2$, we need to find an example
of a regular $G_2$-holonomy manifold that is metrically $\Sigma\times M_4$,
where $\Sigma$ is a supersymmetric 3-cycle on which we wrap an $M2$
brane once, and $M_4$ is a four manifold. One such example was
obtained in \cite{gyz}, which is a $G_2$-holonomy manifold that
is $M_4\times T^3$, metrically. It is given below:
\begin{equation}
\label{eq:G2metric}
ds_7^2= dr^2+{r^2\over4}\sum_{i=1}^3\sigma_i^2 + \sum_{i=1}^3\alpha_i^2,
\end{equation}
where $\sigma_i$'s are left-invariant one-forms obeying the SU(2)
algebra: $d\sigma_i=-{1\over2}\epsilon^{ijk}d\sigma^j\wedge d\sigma^k$,
given by:
\begin{eqnarray}
\label{eq:sigmas}
& & \sigma_1\equiv cos\psi d\theta + sin\psi sin\theta d\phi,
\nonumber\\
& & \sigma_2\equiv -sin\psi d\theta + cos\psi sin\theta d\phi,\nonumber\\
& & \sigma_3\equiv d\psi + cos\theta d\phi,
\end{eqnarray}
and $\alpha_i$'s are harmonic one-forms that form a basis for
$H^1(T^3,{\bf R})$. One can write $\alpha_i=d\theta_i$. To see
that the $T^3$ corresponds to a supersymmetric 3-cycle, we need to 
show that the pull-back of the calibration $\Phi_3$ restricted to $\Sigma$,
is the volume form on $\Sigma$ (See \cite{beckersetal}).
$\Phi_3$ using the notations of \cite{gyz} is given by:
\begin{equation}
\label{eq:caldieuf}
\Phi_3=e^{125} + e^{147} +e^{156}-e^{246} + e^{237}+e^{345} +e^{567},
\end{equation}
where $e^{ijk}\equiv e^i\wedge e^j \wedge e^k$.
Let $1,...,7$ denote $r,\psi,\theta,\phi,\theta_1,\theta_2,\theta_3$.
Hence, when restricted to $\Sigma(\theta_1,\theta_2,\theta_3)$ using
the static gauge, one gets:
\begin{equation}
\label{eq:susySigma}
\Phi_3|_\Sigma=e^{567}=d\theta_1\wedge d\theta_2\wedge d\theta_3,
\end{equation}
which is the volume form on $T^3$. Thus, the $T^3$ of 
(\ref{eq:G2metric}) is a supersymmetric 3-cycle.

For (\ref{eq:G2metric}), one sees  that $g_{ij}=\delta_{ij}+
\partial_i y_0^{\hat{U}}\partial_j y_0^{\hat{V}}g_{\hat{U}\hat{V}}$,
having used the definition of $g_{ij}$ as a pull-back of the
space-time metric $g_{MN}$, static gauge and that $\partial_i y_0^u=0$.
If one assumes that the coordinates $r,\psi,\theta,\phi$ are
very slowly varying functions of $\theta_1,\theta_2,\theta_3$,
one sees that $g_{ij}\sim\delta_{ij}$. This simplifies the algebra,
though one can work to any desired order 
in $(\partial_i y_0^{\hat{U}})^{p(>0)}$,
and get conclusions similar to the ones obtained below. 

Using the reasoning as given in Appendix A, one gets:
\begin{eqnarray}
\label{eq:SdWO1}
& & e_0(x,{\cal O}_1)=(4\pi)^{-{3\over2}};\nonumber\\
& & e_2(x,{\cal O}_1)=0,
\end{eqnarray}
and:
\begin{eqnarray}
\label{eq:SdWO2}
& & e_0(x,{\cal O}_2)=(4\pi)^{-{3\over2}};\nonumber\\
& & e_2(x,{\cal O}_2)=(4\pi)^{-{3\over2}}{72\alpha_2\over r^4}.
\end{eqnarray}

We now do a heat-kernel asymptotics analysis of the fermionic determinant
$det{\cal O}_3$.
The fermionic operator ${\cal O}_3$ can be expressed as:
\begin{eqnarray}
\label{eq:O3}
& & {\cal O}_3\equiv \sqrt{h}h^{ij}\Gamma_jD_i=\sqrt{h}h^{ij}\Gamma_j
\biggl(\partial_i+{1\over4}\omega^{a^\prime b^\prime}_i
\Gamma_{a^\prime b^\prime}
+{1\over4}\omega^{a^\prime b^{\prime\prime}}_i\Gamma_{a^\prime b^{\prime\prime}}\biggr)
\nonumber\\
& & \equiv G^{ij}\Gamma_j\partial_i-r,
\end{eqnarray}
where
\begin{eqnarray}
\label{eq:Grdieufs}
& & 
G^{ij}\equiv\sqrt{h}h^{ij};\nonumber\\
& & r\equiv{-1\over4}\sqrt{h}h^{ij}\Gamma_j\biggl(
\omega^{a^\prime b^\prime}_i\Gamma_{a^\prime
b^\prime}
+\omega^{a^\prime b^{\prime\prime}}_i\Gamma_{a^\prime b^{\prime\prime}}\biggr).
\end{eqnarray}
${\cal O}_3$ is of the Dirac-type as ${\cal O}_3^2$ is of the Laplace-type,
as can be seen from the following:
\begin{eqnarray}
\label{eq:O3squared}
& & {\cal O}_3^2\equiv G^{ij}\partial_i\partial_j+A^i\partial_i+B,\ {\rm where}:
\nonumber\\
& & G^{ij}\equiv hh^{ij};\nonumber\\
& & A^i\equiv\sqrt{h}\Gamma^j\Gamma_k\partial_j(\sqrt{h}h^{kl})
+2h\Gamma^i\Gamma^l\omega_l^{CD}\Gamma_{CD};\nonumber\\
& & B\equiv\sqrt{h}\Gamma^j\Gamma_k\Gamma_{CD}\partial_j(\sqrt{h}h^{kl}\omega_l^{CD})
+h\Gamma^j\Gamma^l\omega_j^{AB}\Gamma_{AB}\omega_l^{CD}\Gamma_{CD}.
\end{eqnarray}
Now, 
\begin{equation}
\label{eq:phi1}
{\cal O}_3\equiv 
G^{ij}\Gamma_j\bigtriangledown_i -\phi
,
\end{equation}
where
\begin{equation}
\label{eq:phi2}
\phi\equiv r+\Gamma^i\omega_i,
\end{equation}
and
\begin{eqnarray}
\label{eq:omegacondieuf}
& & \omega_l\equiv{G_{il}\over2}(-\Gamma^j\partial_j\Gamma^i+\{r,\Gamma^i\}+G^{jk}\Gamma^i_{jk})
\nonumber\\
& & ={h_{il}\over2\sqrt{h}}\biggl({1\over4}\sqrt{h}h^{i^\prime j^\prime}
\{\Gamma_{j^\prime}(\omega^{a^\prime b^\prime}_{i^\prime}\Gamma_{a^\prime
b^\prime}
+\omega^{a^\prime b^{\prime\prime}}_{i^\prime}
\Gamma_{a^\prime b^{\prime\prime}}),
\Gamma^i\}\nonumber\\
& & 
+{hh^{jk}h^{ii^\prime}
\over2}\biggl(\partial_j\biggl[{h_{ki^\prime}\over\sqrt{h}}\biggr]
+\partial_k\biggl[{h_{ji^\prime}\over\sqrt{h}}\biggr]
-\partial_{i^\prime}\biggl[
{h_{jk}\over\sqrt{h}}\biggr]\biggr)\biggr).
\end{eqnarray}

The Seeley-de Witt coefficients $a_i$ are given by (See \cite{Gilkey2}):
\begin{eqnarray}
\label{eq:fermionais}
& & a_1(x,G^{ij}\Gamma_j\bigtriangledown_i -\phi)=-(4\pi)^{-{3\over2}}tr(\phi);\nonumber\\
& & a_3(x, G^{ij}\Gamma_j\bigtriangledown_i -\phi)=-{1\over6}(4\pi)^{-{3\over2}}tr(
\phi\tau+6\phi{\cal E}-\Omega_{a^\prime b^\prime;a^\prime}
\Gamma_{b^\prime}),
\end{eqnarray}
where
\begin{equation}
\label{eq:cal Edieuf}
{\cal E}\equiv-{1\over2}\Gamma^i\Gamma^j\Omega_{ij}+\Gamma^i\phi_{;i}-\phi^2,
\end{equation}
and
\begin{equation}
\label{eq:Omegadieuf}
\Omega_{ij}\equiv\partial_i\omega_j-\partial_j\omega_i+[\omega_i,\omega_j].
\end{equation}
Now, e.g., $\Gamma^i=\partial^i y^M\Gamma_M$, 
where $y^M\equiv y^{m^\prime,\hat{U},u}$
and given that $\partial_i y^u=0$, then in the static gauge,
$\Gamma^i=\delta^i_{m^\prime}\Gamma_{m^\prime}+\partial^i y^{\hat{U}}
\Gamma_{\hat{U}}$
$=\delta^i_{m^\prime}e_{m^\prime}^{\ a^\prime}\Gamma_{a^\prime}$
+ $\partial^i y^{\hat{U}} e_{\hat{U}}^{\ \hat{A}}\Gamma_{\hat{A}}$. Now, 
lets make the simplifying
assumption as done for the bosonic operators, we assume that 
$y^{\hat{U}}$ varies very slowly w.r.t. the $M2$-brane world-volume coordinates.
Hence, we drop all terms of the type $(\partial_i y^{\hat{U}})^{p(>0)}$.
The conclusion below regarding the vanishing of the Seeley-de Witt
coefficients $a_1$ and $a_3$, will still be valid by generalizing
(\ref{eq:trgammas}). The dropping of $(\partial_i y^U)^{p(>0)}$-type terms
will be indicated by $\sim$ as opposed to $=$ in the equations below.

Let $(m,n)$ denote the number of ($\Gamma_{a^\prime}$'s,
$\Gamma_{a^{\prime\prime}}$'s).
Then, without taking into account the multiplicity of the different types of terms,
\begin{eqnarray}
\label{eq:mns}
& & \phi\sim(odd,even)+(even,odd);\nonumber\\
& & \omega_i\sim(even,even)+(odd,odd);\nonumber\\
& & \phi_{;i}\sim(odd,even)+(even,odd),\nonumber\\
& & 
\end{eqnarray}
where $;$ denotes covariant differentiation, using which one gets
\begin{eqnarray}
\label{eq:ai's1}
& & \phi^3\sim(odd,even)+(even,odd);\nonumber\\
& & \phi\Gamma^i\Gamma^j\Omega_{ij}\sim(odd,even)+(even,odd);\nonumber\\
& & \Omega_{a^\prime b^\prime;b^\prime}\Gamma_{a^\prime}\sim(odd,even)+(even,odd).
\end{eqnarray}
Using (\ref{eq:D=11Gammadefs}), one sees that
\begin{equation}
\label{eq:trgammas}
tr(\prod_{i=1}^{2m+1}\Gamma_{a^\prime_i})
=tr(\prod_{i=1}^{2m}\Gamma_{a^\prime_i}\Gamma_{a^{\prime\prime}})=
tr(\prod_{i=1}^{2m+1}
\Gamma_{a_i^\prime}\Gamma_{a^{\prime\prime}b^{\prime\prime}})=
tr(\prod_{i=1}^{2m}
\Gamma_{a_i^\prime}\Gamma_{a^{\prime\prime}}\Gamma_{b^{\prime\prime}}
\Gamma_{c^{\prime\prime}})=0
\end{equation}
Hence,
\begin{equation}
\label{eq:SdWferm1}
a_1(x, G^{ij}\Gamma_j\bigtriangledown_i -\phi)
=a_3(x, G^{ij}\Gamma_j\bigtriangledown_i -\phi) \sim0.
\end{equation}
We conjecture that in fact, $a_{2n+1}(x,G^{ij}\Gamma_j\bigtriangledown_i-\phi)\sim0$
for $n=0,1,2,3,...$.

By using reasoning similar to the one used in Appendix A, one can show that:
\begin{eqnarray}
\label{eq:O3sqzetaSdW}
& & e_0(x,{\cal O}_3^2)=(4\pi)^{-{3\over2}};\nonumber\\
& & e_2(x,{\cal O}_3^2)\sim0.
\end{eqnarray}
From the extra factor of ${1\over2}$ multiplying the 
$\zeta^\prime(0|{\cal O}_3^2)$ relative to $\zeta^\prime(0|{\cal O}_1)$ in
(\ref{eq:bosfermdets}), and (\ref{eq:SdWferm1}) and (\ref{eq:O3sqzetaSdW}),
one sees the possibility that:
\begin{equation}
\label{eq:bosfermcan}
{det {\cal O}_3\over det {\cal O}_1}\sim{1\over2}.
\end{equation}

In conclusion, one sees that Seeley-de Witt coefficients of the
fermionic operator ${\cal O}_3$ are proportional to  those of  
the bosonic operator
 ${\cal O}_1$ in the adiabatic approximation, to the order calculated,
 for the $G_2$-metric 
(\ref{eq:G2metric}). This is indicative of
possible cancellation between them. This is expected, as the $M2$-brane action
has some supersymmetry. As $b_1(T^3)=3>0$, thus the supersymmetric
3-cycle of (\ref{eq:G2metric}) is an example of a non-rigid 
supersymmetric 3-cycle. The result of \cite{harvmoore} is not applicable
for this case. On the other hand, the superpotential written out
as determinants, as in this work, is still valid.
The corresponding modified formula in \cite{harvmoore} might consist, as 
prefactors, in addition to the phase, the torsion elements of
$H_1(\Sigma,{\bf Z})$, represented by $|H_1(\Sigma,{\bf Z})|^\prime$
in \cite{RozWitt}, and perhaps a geometrical quantity that would encode
 the $G_2$-analog of the arithmetic genus condition 
in the context of 5-brane instantons obtained by wrapping $M5$-brane
on supersymmetric 6-cycles in $CY_4$ in \cite{wittennonpertW}
\footnote{We are grateful to S.Gukov for telling us about the
topological 
condition involving the arithmetic genus in \cite{wittennonpertW}.}.
The validity of the arithmetic genus argument, even for $CY_4$,
 needs to be independently verified though\cite{wittencom}.
The above cancellation implies that for the $G_2$-metric
(\ref{eq:G2metric}), the geometric prefactors will be $\sim
\sqrt{{1\over2\det{\cal O}_2}}$. This
could perhaps be a generic feature with all $G_2$-metrics that are metrically
$\Sigma\times M_4$. 

\section{Conclusion and Discussions}

In this paper, we have evaluated in a closed form, the exact expression for the
nonperturbative contribution to the superpotential from a single $M2$-brane 
wrapping an isolated supersymmetric 3-cycle of a $G_2$-holonomy manifold. 
The comparison with Harvey and Moore's result, is 
suggestive but not exact. A heat-kernel asymptotics analysis
for a non-compact smooth $G_2$-holonomy manifold that is metrically
${\bf R}^4\times T^3$, in the adiabatic approximation, showed that
the UV-divergent terms of one of the bosonic and the fermionic determinants
are proportional to each other, to the order we calculate, 
indicative of cancellation between the same, as expected because
the $M2$ brane action of Bergshoeff, Sezgin and Townsend
is supersymmetric. {\it Unlike the result of \cite{harvmoore},
the expression obtained for the superpotential above in terms of
fermionic and bosonic determinants, in addition to a holomorphic phase 
factor, is valid even for non-rigid supersymmetric 3-cycles as the
one considered in (\ref{eq:G2metric}) above.}

Following \cite{gubseretal}, it is 
tempting to conjecture that the superpotential term corresponding to
multiple wrappings of the $M2$-brane around the supersymmetric 3-cycle, should be give by:
\begin{equation}
\label{eq:mults}
\Delta W=\sum_n\sqrt{{det{\cal O}_3\over\det{\cal O}_1det{\cal O}_2}}{e^{n\int_\Sigma[iC-
{1\over l_{11}^3} vol(h)]}\over n^2}.
\end{equation}
The calculations in this paper were done for a smooth $G_2$-holonomy manifold. However,
for singular $G_2$-holonomy manifolds with $ADE$-type singularity, e.g., spin bundles over
$S^3$ with the topology ${{\bf C}^2\over\Gamma_{ADE}}\times S^3$, one can explore 
the following in more details. Based on \cite{SWHollow}, wherein the four dimensional
superpotential of Super Yang-Mills theory, is calculated by compactifying the theory
on a circle, and then taking its four dimensional limit, it was suggested in 
\cite{acharya} that the nonperturbative superpotential of membrane instantons in $M$-theory
on ${{\bf C}^2\over\Gamma_{ADE}}\times S^3\times S^1\times{\bf R}^{2,1}$ 
can be evaluated by compactifying 
$M$-theory on the $S^1$ to type $IIA$ theory on 
${{\bf C}^2\over\Gamma_{ADE}}\times S^3\times{\bf R}^{2,1}$ and evaluating the 
superpotential from the wrapping of $D2$ branes on the 
(supersymmetric) $S^3$, whose
world-volume theory in the large volume limit of the $S^3$, is given by a ``quiver gauge
theory" \cite{dougmoore}. The wrapped $D2$ branes are constrained to live at
the singularities (that lie at the origin of ${{\bf C}^2\over\Gamma_{ADE}}$) of
the $G_2$-holonomy manifold, which in the quiver gauge theory, is the 
locus in the moduli space, where the $D2$ branes fractionate. These 
fractional $D2$ branes are supposed to be $D4$ branes wrapping vanishing 2-cycles at
the origin in ${{\bf C}^2\over\Gamma_{ADE}}$.

In terms of relating the result obtained in (\ref{eq:Wfinal}) to that of
the 1-loop Schwinger computation of $M$ theory and the large $N$-limit of the
partition function evaluated in \cite{GopVafaMth1}\footnote{This
logic was suggested to us by R.Gopakumar.}, one notes that
the 1-loop Schwinger computation also has as its starting point, an 
infinite dimensional
bosonic determinant of the type $det\biggl((i\partial-e A)^2
-Z^2\biggr)$, $A$ being the gauge field
corresponding to an external self-dual field strength, and $Z$ denoting the 
central charge. The large $N$-limit of the partition function of Chern Simons theory
on an $S^3$, as first given by Periwal in \cite{Periwal}, involves the product of infinite
number of $sin$'s, that can be treated as the eigenvalues of an infinite determinant. 
This is indicative 
of a possible connection between the membrane instanton contribution
to the superpotential, the 1-loop Schwinger computation and the large $N$ limit
of the Chern-Simons theory on an $S^3$.

\section*{Acknowledgements}

We would first like to thank R.Gopakumar for introduction to and
having innumerable discussions on issues related to open/closed string
dualities, for bringing to our notice and discussing \cite{harvmoore},
and for a critical reading of the manuscript.
We are grateful to D.Jatkar, and especially S.Gukov,  for several clarifications
(thanks to S.Gukov for bringing \cite{wittennonpertW} to our notice)
and D.Ghoshal for suggesting \cite{Gilkey}.
Its a pleasure to thank K.Kirsten, and P.Gilkey, in particular,
for a number of clarifications on heat kernel techniques, and
for bringing \cite{Deseretal,Gilkey2} to our attention. 
We are grateful to J.Park for going through a 
preliminary version of Section {\bf 2} and making useful 
comments. We thank E.Witten for a clarification regarding membranes wrapped
around non-rigid supersymmetric divisors.
We learnt about \cite{dougmoore,SWHollow,acharya}
from B.S.Acharya's lectures at the spring school on ``Superstrings and
Related Matters" at the Abdus Salam I.C.T.P., Trieste, Italy, from
Mar 18-26, 2002.

\appendix
\section*{Seely -de Witt coefficients for the bosonic determinants
in the superpotential}
\setcounter{equation}{0}
\seceqaa

In this  appendix we derive (\ref{eq:SdWO1}) and (\ref{eq:SdWO2}).

Lets first consider the Seeley de-Wit coefficients for ${\cal O}_1$.
Now, in the above adiabatic approximation, the world volume metric of
the $M2$-brane is flat. Hence, the Christoffel connection $\Gamma^i_{jk}$
in (\ref{eq:omegadieuf}) for ${\cal O}_1$, vanishes. Now, 
$\omega^{a^\prime b^\prime}_i\sim\delta^{m^\prime}_i
\omega_{m^\prime}^{a^\prime b^\prime}$, where 
\begin{equation}
\label{eq:spcon}
\omega_{m^\prime}^{ab} = e^{[a}_{\ n^\prime}g^{n^\prime l^\prime}
(\partial_{m^\prime} e_{l^\prime}^{\ b^\prime]} - 
\Gamma^{p^\prime}_{l^\prime m^\prime}e_{p^\prime}^{\ b^\prime]}),
\end{equation}
the antisymmetry indicated on the right hand side of 
(\ref{eq:spcon}) being applicable only to the tangent-space indices
$a^\prime,\ b^\prime$, and 
where for (\ref{eq:G2metric}), the following are the non-zero vielbeins:
\begin{eqnarray}
\label{eq:viels}
& & e^1_{\ r}=1;\nonumber\\
& & e^2_{\ \theta}={r\over2}cos\psi,\ e^2_{\ \phi}={r\over2}sin\psi sin\theta;
\nonumber\\
& & e^3_{\ \theta}=-{r\over2}sin\psi,\ e^3_{\ \phi}=cos\psi sin\theta;
\nonumber\\
& & e^4_{\ \psi}={r\over2},\ e^4_{\ \phi}={r\over2}cos\theta;\nonumber\\
& & e^5_{\ \theta_1}=e^6_{\ \theta_2}=e^7_{\ \theta_3}=1.
\end{eqnarray}
Hence, for the $G_2$ metric of (\ref{eq:G2metric}), 
$\partial_{m^\prime}e_{l^\prime}^{\ b^\prime}=0$. Also, 
$\Gamma^{p^\prime}_{l^\prime m^\prime}=0$. Thus, $\omega^{\cal D}_i\sim 0$.

In the adiabatic approximation, $\tau\sim0$. 

Hence, 
\begin{eqnarray}
& & e_0(x,{\cal O}_1)=(4\pi)^{-{3\over2}};\nonumber\\
& & e_2(x,{\cal O}_1)=0.
\end{eqnarray}

We next consider evaluation of $e_{0,2}(x,{\cal O}_2)$. Once again,
the Christoffel connection $\Gamma^i_{jk}$ vanishes. Again, 
$\omega^{D,{\cal D}}\sim 0$. Also, $\partial_i h_{\hat{U}\hat{V}}\sim0$.
Hence, $A^i\sim 0$ that figures in (\ref{eq:A^iBdieufs}).

Now,
\begin{equation}
\label{eq:Udieuf}
{\cal U}_{\hat{U}\hat{V}}\equiv{1\over2} R^{m^\prime}_{\hat{U}m^\prime\hat{V}}
+{1\over8}Q^{m^\prime n^\prime}_{\hat{U}}Q_{m^\prime n^\prime\hat{V}},
\end{equation}
where the second fundamental form is defined via:
\begin{equation}
\label{eq:Qdieuf}
\Gamma^{m^{\prime\prime}}_{k^\prime l^\prime}\equiv -{1\over2}
g^{m^{\prime\prime}n^{\prime\prime}}
Q_{k^\prime l^\prime n^{\prime\prime}}.
\end{equation}
Using:
\begin{equation}
\label{eq:Rdieuf}
R^{m^\prime}_{\hat{U}m^\prime\hat{V}}=
\partial_{\hat{V}}\Gamma^{m^\prime}_{\hat{U}m^\prime}
-\partial_{m^\prime}\Gamma^{m^\prime}_{\hat{U}\hat{V}}
+\Gamma^V_{\hat{U}m^\prime}\Gamma^{m^\prime}_{V\hat{V}}
-\Gamma^V_{\hat{U}\hat{V}}\Gamma^{m^\prime}_{Vm^\prime},
\end{equation}
and the fact that the non-zero Christoffel symbols do not involve
$m^\prime$ as one of the (three) indices and that their values are
$m^\prime$-independent, one sees that 
\begin{equation}
\label{eq:nullU}
{\cal U}_{\hat{U}\hat{V}}=0.
\end{equation}
Hence, $B\sim0$ that figures in (\ref{eq:A^iBdieufs}). 

For evaluating $\tau\equiv g^{i_1 i_2}g^{j_1 j_2}R_{i_1 j_1 j_2 i_2}$
$=g^{i_1 i_2}g^{j_1 j_2}g_{i_1 l_1}R^{l_1}_{j_1 j_2 i_2}$, one
needs to evaluate $R^{l_1}_{j_1 j_2 i_2}=\partial_{i_2}\Gamma^{l_1}_{j_1 j_2}$
-$\partial_{j_2}\Gamma^{l_1}_{j_1 i_2}$ $+\Gamma^p_{j_1 j_2}\Gamma^l_{p i_2}$
-$\Gamma^p_{j_1 i_2}\Gamma^{l_1}_{p j_2}$. This will be evaluated using
the metric given by $G^{ij}=g^{ij}\sqrt{g}h_{\hat{U}
\hat{V}}$, as given in (\ref{eq:O12qts}), 
where we will use the adiabatic approximation:
$g_{ij}\sim\delta^{m^\prime}_i\delta^{n^\prime}_jg_{m^\prime n^\prime}$.
Due to the $\hat{U}\hat{V}$ indices, the Ricci scalar $\tau$ is 
actually a matrix in the $X_{G_2}\setminus\Sigma$ space.
In the adiabatic approximation, only the product of the two 
Christoffel symbols in $R^{l_1}_{j_1 j_2 i_2}$ is non-zero, and is
given by the following expression:
\begin{eqnarray}
\label{eq:tauexp}
& & \Gamma^{p^{\prime\prime}}_{j_1 j_2}\Gamma_{p^{\prime\prime}i_2}^{l_1}
-\Gamma^{p^{\prime\prime}}_{j_2 i_1}\Gamma^{l_1}_{p^{\prime\prime} j_2}
\nonumber\\
& & =-{h_{\hat{U}\hat{V}}\over4}\delta_{j_1 j_2}\delta^{l_1}_{i_2}\Biggl[
\biggl(\partial_r\biggl[{1\over h_{\hat{U}\hat{V}}}\biggr]\biggr)^2
+{4\over r^2}\biggl(\partial_\theta\biggl[{1\over h_{\hat{U}\hat{V}}}\biggr]
\biggr)^2\Biggr]\nonumber\\
& & =-{\delta_{j_1 j_2}\delta^{l_1}_{i_2}\over 4}
\left(\begin{array}{cccc}\\
0 & 0 & 0 & 0 \\
0 & {16\over r^4} & 0 & {16\over r^4 cos\theta}
+{4sin^2\theta\over r^2 cos^3\theta} \\
0 & 0 & {16\over r^4} & 0 \\
0 & {16\over r^4 cos\theta}+{4sin^2\theta\over r^2 cos^3\theta} 
& 0 & {16\over r^4} \\
\end{array}\right).
\end{eqnarray} 
Hence, on taking $tr_{X_{G_2}\setminus\Sigma}$, one gets:
\begin{equation}
\tau={72\over r^4}.
\end{equation}
Hence,
\begin{eqnarray}
& & e_0(x,{\cal O}_2)=(4\pi)^{-{3\over2}};\nonumber\\
& & e_2(x,{\cal O}_2)=(4\pi)^{-{3\over2}}{72\alpha_2\over r^4}.
\end{eqnarray}

\end{document}